# Non-Hermitian arrangement for stable semiconductor laser arrays

## J. Medina Pardell,[1,*] R. Herrero,[1] M. Botey,[1] K. Staliunas[1,2]

[1] *Departament de Física, Universitat Politècnica de Catalunya (UPC), Colom 11, E-08222 Terrassa, Barcelona, Catalonia;*
[2] *Institució Catalana de Recerca i Estudis Avançats (ICREA), Passeig Lluís Companys 23, E-08010 Barcelona, Catalonia;*
*\*judith.medina@upc.edu*

**Abstract:** We propose and explore a physical mechanism for the stabilization of the complex spatiotemporal dynamics in arrays (bars) of broad area laser diodes taking advantage of the symmetry breaking in non-Hermitian potentials. We show that such stabilization can be achieved by specific pump and index profiles leading to a PT-symmetric coupling between nearest neighboring lasers within the semiconductor bar. A numerical analysis is performed using a complete (2+1)-dimensional space-temporal model, including transverse and longitudinal spatial degrees of freedom and temporal evolution of the electric field and carriers. We show regimes of temporal stabilization and light emission spatial redistribution and enhancement. We also consider a simplified (1+1)-dimensional model for an array of lasers holding the proposed non-Hermitian coupling with a global axisymmetric geometry. We numerically demonstrate a two-fold benefit: the control over the temporal dynamics over the EELs bar and the field concentration on the central lasers leading to a brighter output beam, facilitating a direct coupling to an optical fiber.

## 1. Introduction

Diode lasers systems, either in the form of a single Edge-Emitting Laser (EEL), in the form of arrays of lasers (lasers bars), or as stacks of EEL bars, are replacing other lasers sources due to their compactness, affordable prices and high performance. However, they present a major drawback, the laser spatiotemporal instability and divergence which, particularly, prevents them to efficiently couple to optical fibers. Spatiotemporal complex dynamics is commonly observed in spatially extended, dissipative systems which are driven by an external pump. Board area diode lasers are no exception and are plagued by self-focusing filamentation instabilities and complex dynamical behaviors [1,2,3]. Moreover, in EELs bars, the coupling between neighboring EELs contributes to increase such spatiotemporal instabilities [4,5,6,7]. The consequence is the onset of chaotic and turbulent regimes produced by spatial and temporal instabilities such as the modulation instability and Hopf bifurcations. Yet, different proposals to reduce or to eliminate these instabilities for a single laser relay on external cavities or elements, therefore compromising the compactness of the laser [8,9,10]; this is also the case for bars of EEL lasers [11,12,13,14]. Besides, a complex optical setup made up of: a fast-axis collimator, a slow-axis collimator and a focusing lens, is commonly used to improve efficient coupling into multimode fibers. Usually, the intrinsic turbulent behavior enlarges the smallest possible focal point and limits the coupling to thin monomode fibers [15]. Possible approaches on the regularization of a single semiconductor laser are based on intracavity filtering [16,17] or on spatial gain profile to mitigate semiconductor lasers instabilities by the introduction of intrinsic complex modulations within the laser [18,19, 20]. But obtaining a stable emission from EELs bars remains a longstanding open question, and there is a need for a compact stabilization scheme.

Recently, the interplay between gain and index modulation has emerged as a fruitful new research area in photonics. Initially introduced as a curiosity in quantum mechanics [21], parity-time (PT-) symmetry, found experimental realizations in the field of photonics in artificial materials with spatial distributions of real and complex permittivities, showing the ability of molding the flow of light [22,23,24,25,26]. The attentions to those systems that while being non-conservative could still hold real energy eigenvalues, derives from the unusual, even counter intuitive properties they hold arising from an asymmetric coupling of modes. Beyond the particular class of open non-conservative systems holding PT-symmetry however, there is a larger class of non-conservative as non-Hermitian Hamiltonians [27]. Indeed complex Non-Hermitian photonics has led to technologically accessible novel effects, from transparency and invisibility [28,29] to light transport [30] including various applications in laser science [31,32]. In particular, the new concepts of non-Hermitian photonics have successfully been applied to the control of the dynamics of broad semiconductor lasers [33,34], and arrays of vertical emitting semiconductor lasers or a ring array of semiconductor lasers [35,36].

Our proposal is intended to obtain a stable emission from an array of EELs and the improvement of its beam quality and energy distribution within the laser array. Altogether allows the direct coupling to fiber or optical guide without any optical component that should strongly enhance the coupling efficiency. The light generated in every single semiconductor laser is expected to be spatially redistributed and temporally stabilized via non-Hermitian coupling between neighboring lasers induced by a particular gain (pump) and index modulation (stripes) of the structure. The system is described by a complete (2+1)-dimensional space-temporal model, including transverse and longitudinal spatial directions and temporal evolution of the electric field and carriers.

We first identify the onset of spatiotemporal instabilities for a single laser source, the regime of temporally stable and monomode emission severely restricts the power of the laser source. However, splitting a broad EEL source in an array of stable, thinner lasers with stable emission parameters, is not a solution, since new temporal and synchronization instabilities arise from the coupling between neighboring lasers leading again to irregular spatiotemporal behaviors. Thus, we propose a non-Hermitian asymmetric coupling between EELs within the array for stabilization and redistribution of the light emission. This emission improvement is first demonstrated for the simple two coupled laser system. We determine the stabilization performance as a function of the shift between the pumped region and laser stripes and the distance between the two lasers. Next, we analyze a system formed by three lasers holding a global mirror symmetry to induce an inward coupling in the laser array. In the following, a simplified (1+1)-dimensional model is used to extend the study to a full EEL bar formed by an array of many lasers. The simulations show both temporal stabilization and simultaneous spatial redistribution, i.e. localization, of the generated light.

## 2. Model for semiconductor laser arrays

In order to model the spatial redistribution and temporal stabilization of coupled EEL sources, we use a well-established model including the spatiotemporal evolution of the electromagnetic field and carrier density inside the cavity [20]. EELs are usually described either by stationary models [37] or dynamical models of the mean field [38]. Here, the complete dynamical model is used for the forward and backward fields propagating within the cavity and the carrier density. It was recently used to demonstrate spatial filtering of broad EEL sources [17]. Since the round-trip time of the cavity (on the order of ps) is small compared to the carrier's relaxation time (on the order of ns), the temporal evolution of the field in one roundtrip may be calculated by its propagation along the cavity assuming constant carriers. Applying the slowly varying envelope approximation, the forward and backward envelopes of the electric field, $A^{\pm}$ are integrated along the EEL followed by the second step, the temporal integration of carriers considering a constant field. We neglect the frequency dependence of material gain, spatial hole burning of carriers, and heating-induced changes of model parameters since we assume we do not reach high powers [39,40]. Overall, this results in the following non-linear system of three coupled equations:

$$\pm \frac{\partial A^{\pm}}{\partial z} = \frac{i}{2k_0 n} \frac{\partial^2 A^{\pm}}{\partial x^2} + \sigma[(1-ih)N - (1+\alpha)]A^{\pm} + i\Delta n(x)k_0 A^{\pm} \quad (1)$$

$$\frac{\partial N}{\partial t} = \gamma(-N - (N-1)(|A^+|^2 + |A^-|^2)^2 + p_0 + \Delta p(x) + D\nabla^2 N)$$

where $k_0$ is the wavevector, $n$ is the effective refractive index, $\sigma$ is a parameter inversely proportional to the light matter interaction length, $h$ is the Henry factor or linewidth enhancement factor of the semiconductor, $\alpha$ corresponds to losses, $p_0$ is the pump, $D$ is the carrier diffusion coefficient and $\gamma$ is the inverse of carriers' relaxation time, $\tau_{nr}$. In our calculations the transverse and longitudinal spatial coordinates are in units of the wavelength, time is normalized to the roundtrip time; $N$ is normalized to $N_0$ (the carrier's density to achieve transparency) and the electric field envelope is normalized to $\frac{a\tau_{nr}}{\hbar\omega}$ being $a$ the gain parameter, $\omega$ the angular frequency of light. Polarization of the material is eliminated in eq. (1) as the semiconductor laser is considered a class B laser and the fine longitudinal interference between the forward and backward fields, considered to be blurred by the carrier diffusion, is disregarded. Finally, the transverse modulations of the refractive index $\Delta n(x)$ account for the individual laser stripes; and the pump, $\Delta p(x-s)$, where $s$ is a spatial shift the spatial profile of the electrodes. Both modulations, $\Delta n(x)$ and $\Delta p(x-s)$, induce the real and imaginary parts of the non-Hermitian potential, which, properly designed, may lead to an asymmetric field coupling. See Fig. 1. (a) for a schematic representation of the laser architecture. In order to avoid discontinuities in the derivatives of these modulations, the two spatial transverse profiles are mathematically described as consecutive sharp sigmoids.

In the proposed scheme, the two profiles, $\Delta n(x)$ and $\Delta p(x)$, can be slightly spatially shifted a distance $s$, one with respect to another, and it is precisely this interplay between index and gain profiles that is expected to induce a non-Hermitian potential and asymmetric coupling between neighboring lasers. The boundary conditions are straightforwardly determined by the Fabry-Perot cavity mirrors located at $z = 0$ and $z = L$ are $A^+(z=0) = r_0 A^-(z=0)$ and $A^-(z=L) = r_L A^+(z=L)$, where $L$ is the length of the laser and $r_{0/L}$ are the corresponding reflection of the edge mirrors at $z = 0/L$, respectively.

First, we numerically study the spatiotemporal behavior of a single EEL source through the system model in eq. (1) to determine its dynamics for different working conditions. As it is well known, decreasing the laser width acts as a mode selection mechanism when light is confined and a the broad and strongly multimode semiconductor emission turns into a monomode emission regime. However, to achieve a brighter source it is not enough to split a broad EEL source into an array of spatially stable thin EELs by patterning longitudinal separation slits between them. It is also necessary to engineer the coupling between lasers within the array to obtain stability and improve the quality of the emission, see Fig. 1(a). Numerical simulations supporting this idea are provided in Fig 1. (b), showing the total emitted output power of a single EEL for widths, $w$, ranging from 2.5 μm to 50 μm, and analyzing the field profile within the laser while decreasing width seeking for the onset of the monomode emission. The maximum width for a monomode emission determined by

$w_{max} = \lambda/\sqrt{2n\Delta n}$ (corresponding to 4 µm for the general operational parameters chosen in Fig.1) is in good agreement with the integrated beam profiles. In turn, we calculate the beam quality factor, $M^2$ as the ratio between the Beam Parameter Product (*BPP*) of a real beam and a Gaussian beam, which can be numerically evaluated as: $M^2 = \frac{BPP_{actual\ beam}}{BPP_{Gaussian\ beam}} = \frac{\pi w \theta}{\lambda}$, where $w$ is the near-field width, and where the divergence, $\theta$, is obtained from the far field, provided in the inset of Fig. 1 (b). Indeed, the transvers cuts for the laser profiles provided in Fig. 1. (b), also evidence how $M^2$ strongly increases with the laser's width as the laser becomes more multimode and inhomogeneous, see However, while a short width in a semiconductor laser indeed acts as a transverse mode selection mechanism it does not warrant its temporal stability. This can be observed in Fig. 1. (c), where the temporal stability of a monomode laser —namely, $w = 2.5$ µm— is scanned as a function of the pump. The results show a Hopf bifurcation arising at a particular pump, referred as the Hopf pump, $p_H$, and for pump values above this threshold the laser becomes temporally unstable as it is evident comparing the temporal evolution of the spatial distribution transverse profile for two given pump values below and above $p_H$, sharing however monomode spatial distribution along the laser. The smooth oscillations observed just after the Hopf bifurcation present periods about 20 roundtrips and are only visible in the small pump interval shown in the figure, while for larger pump values sharp peaks with an almost zero background appear. Scanning the pump upwards, pulses rapidly increase in amplitude while becoming shorter, with a time duration of few roundtrips, and reducing their frequency.

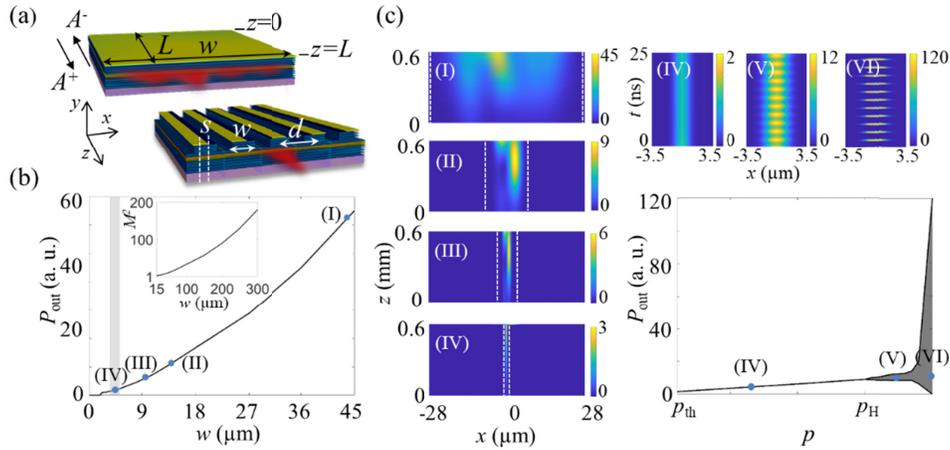

**Fig. 1.** (a) Scheme of a broad single EEL where the index and pump profiles coincide with no spatial shift ($s = 0$µm), where $w$ is the laser width and $L$ the length, and scheme of a laser array made of single lasers with $s \neq 0$, separated a distance $d$. (b) Calculated output of the EEL as a function of the laser width, as obtained from eq. (1); inset: $M^2$ dependence on the laser width for a given pump $p_0 = 2.5$. The lateral subplots depict the spatial intensity distributions for different widths of the pump: I 50 µm, II 12.5µm, III 7.5µm, IV 2.5µm determined by dashed lines on each plot; grey area: analytical range of monomode emission. (c) Temporal evolution of the transverse intensity profile within the laser for $w = 2.5$ µm and different pumps, normalized to the emission threshold $p_{th}$: IV $p = 1.1\ p_{th}$, V $p = 1.8\ p_{th}$, VI $p = 1.9\ p_{th}$, and numerical Light-Intensity output power, LI-curve, as a function of the pump; $p_H$ indicates the onset of the instability. The integration parameters are: $L = 500$, $\alpha = 0.1$, $h = 2.0$, $\sigma = 0.06$, $D = 0.03$ and $\gamma = 0.005$, and the units in the graphs correspond to $\lambda = 1$µm, with $n_0 = 3.5$ and $\Delta n = 0.06$.

## 3. Symmetric and asymmetric coupling

Once the main dynamics and parameters of a single laser are determined we proceed to analyze the effect of the coupling between lasers on the dynamics. Such coupling depends on the distance between neighboring lasers and can be further engineered by introducing a displacement between the laser profile and the pump, as schematically shown in Fig. 1. (a). We first analyze the coupling between two identical lasers with the same intrinsic parameters and where the index and the pump profiles perfectly coincide, and are, therefore, symmetrically coupled. While the two standing alone lasers may have a spatially and temporally stable emission, as the distance between them decreases —the coupling strength increases— and keeping the rest of parameters, both lasers become temporally unstable. Spatial asymmetries are evident in every snapshot of the numerically calculated intensity distribution of two close EEL sources, see Fig. 2. (a). Besides, the temporal evolution of the transverse profile of the intensity at any position of the cavity length is aperiodic, see Fig. 2. (b). Next, we slightly shift the index profile of both lasers with respect to the gain profiles to induce a mirror-symmetric coupling. As expected, the light generated in one laser is partially transferred to the other one, see Fig. 2. (c). As an important consequence, when the energy is redistributed due to the asymmetric coupling, both lasers become temporally stable, as shown in the temporal evolution of the intensity transverse profile in Fig. 2. (d). This temporal stabilization tendency is in agreement with the general behavior of coupled nonlinear oscillators generally showing less complex dynamics for unidirectional than for bidirectional couplings.

The performance of the proposed asymmetric coupling is assessed by the asymmetric energy enhancement and temporal stability of the attained regimes. We calculate the enhancement as the relative intensity of the laser to which the energy is accumulated, i.e. as the ratio of the temporally averaged intensity of the enhanced laser, for $s \neq 0$, versus the unshifted case, $s = 0$. We explore the parameter space of the distance between lasers, $d$, and asymmetry shift parameter, $s$ (spatial shift between the pump and the refractive index profiles) for a fixed value of the pump. The results are summarized in Figs. 2. (e). While a larger enhancement could be expected by increasing the shift parameter $s$. However, we observe that the emission decreases for a maximum around a given shift value, namely $s \approx 0.25$ µm. This decrease may be attributed to the asymmetric configuration of both lasers which also induces an asymmetric leaking of energy opposite to the direction of the other laser. Such leaked energy is therefore lost, and the net gain for the whole system is reduced; being the largest relative intensity of the enhanced factor around about 2. Moreover, such enhancement increases for a smaller distance, $d$. In addition, the temporal stability of the emission may be evaluated by mapping the amplitude of the temporal oscillations of the enhanced laser also in the distance-shift, $(d,s)$, parameter space. Temporal instabilities arise for small $d$ and $s$ values, i.e. when the distance between lasers or coupling asymmetry decrease. For center-to-center distances close to the laser width the emission is found to be unstable for all values of $s$. On the contrary, stability is found either increasing the coupling asymmetry for a given laser distance, as also, trivially, at larger distances between lasers. Interestingly, inspecting Figs. 2. (e) and (f), we observe that there is a range of parameters around the maximum relative intensity region which coincides with a temporally stable behavior.

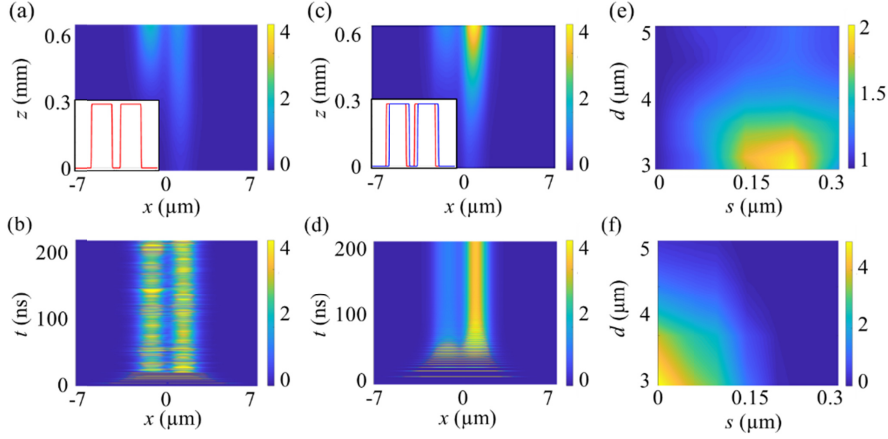

**Fig. 2.** (a)/(c) Snapshots of the spatial intensity distribution after sufficient integration time for two symmetrically/asymmetrically coupled lasers, and (b)/(d) temporal evolution of the intensity at $z = L$ for two symmetrically/asymmetrically coupled lasers with $w = 2.5 \mu m$, separated a center-to-center distance $d = 3$ μm and for $s = 0.25$ μm. Insets show the transverse pump profile $\Delta p$ (red curve) and the index profile $\Delta n$ (blue curve), and they coincide in (a). (e) Intensity enhancement factor of the amplified laser depending on the distance between lasers, $d$, and spatial shift, $s$. (f) Temporal instability map. Relative amplitude of the temporal oscillations (normalized to the average amplitude) All the integration parameters are the same as in Fig. 1, and $p_0 = 2.5$.

As above mentioned, while the energy is enhanced in one laser, some energy is in turn leaked due to the asymmetric configuration, asymmetrically coupled out from the laser. We further analyze this effect considering a single laser. We observe that around $s = 0.35$ μm all the energy created in the laser is lost, see Fig. 3. (a). A physical insight to this effect is found calculating the transverse spatial tilt of the phase of the electric field within the laser that indicates a transverse shift of light propagating along the laser and out from the laser stripe, see Fig. 3. (b). As it may be expected, such phase if completely symmetric for $s=0$, while the averaged phase slope inside the laser, directly increases with the asymmetry parameter $s$, see Fig. 3. (c), indicating a linear increase of the transverse transfer of energy with $s$.

Therefore, the effect is optimized by considering a single transverse shift, i.e. no shift for the laser to which the energy is transferred. We compare this last configuration with the double transverse shift configuration inspected in Fig. 2 for a distance of 3 μm between the two lasers, see Fig.3 (d). The double shifted configuration achieves a maximum of output power for an $s$ value about 0.25 μm and decrease due to the energy leakage while the single shifted case keeps increasing with $s$. Temporal instabilities also disappear for the single shift case increasing $s$ parameter and the amplitude of the temporal oscillations shows a similar $(d, s)$ dependence of Fig. 2. f. The spatial intensity distribution and the temporal evolution for the single shift configuration are provided in Figs. 3. (e) and (f) showing more than twice the emission of the double shifted scheme of Fig. 2. (c).

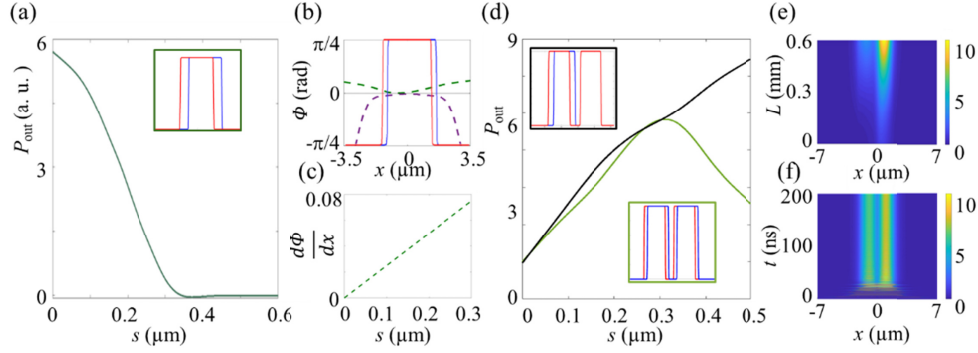

**Fig. 3.** (a) Loss of output power for one laser when $s$ increases, being $s$ the transverse shift between $\Delta n$ and $\Delta p$, in blue and red in the inset. (b) Field phase for $s = 0$ μm (dashed purple curve) and for $s = 0.15$ μm (dashed green curve). The blue and red solid lines represent $\Delta n$ and $\Delta p$ for the second case. (c) Field phase slope at the central point of the laser as a function of $s$. (d) Comparison of the output power for two coupled lasers with asymmetric lateral coupling with single (black) and double transverse shift (green) as a function of $s$. (e) Snapshot of the spatial intensity distribution after sufficient integration time and (f) temporal evolution of the intensity for two coupled lasers with single transverse shift corresponding to the black curve for $s = 0.4$ μm. All the integration parameters are the same as in Fig. 2.

## 4. Axisymmetric inward coupling of EELs

As observed, the asymmetric coupling leads to simultaneous enhancement and temporal stabilization of the emitted field and the effect is optimized when the enhanced laser has no asymmetric shift. Therefore, the next step is considering a simple array architecture formed by three coupled lasers with asymmetric coupling but with a central symmetry axis [33]. That is to say, two symmetric lasers holding asymmetric shift between gain and index profiles and a central one with no shift. For such configuration, the coupling is expected to increase as the distance decreases eventually also reaching a temporally stable regime as light is localized inwards.

The numerical results for the three coupled lasers are summarized on Figs. 4. It can be observed that when symmetrically coupled, all three lasers are temporally unstable as intensity switches from one laser to another see Figs. 4 (a) and (b). On the contrary, as the non-Hermitian inward potential is introduced, by means of the inward asymmetric shift of the index profile versus the gain profile, the generated light tends towards the central laser and all lasers become stable, see Fig. 4. (c) and (d). We note that in this scenario, the fulfilled redistribution and spatial stabilization leads to enhanced intensities as compared with the two-laser case.

The performance of the proposed mirror symmetric lasers triad is evaluated by the relative intensity of the central laser, i.e. intensity of the enhanced laser, for $s \neq 0$, versus the unshifted case, $s = 0$. Comparing Fig. 4. (e) and Fig. 3. (e), we observe the maximum intensity for a somewhat larger value of $d$. Besides, a larger enhancement ratio is reached, as energy is coupled from the two neighboring lasers. The temporal stability of the emission is here evaluated by mapping the temporal oscillations of the central laser also in the distance-shift parameter space, $(d, s)$, for a fixed pump value. The results, summarized in Fig. 4. (f), show a range of maximum energy localization, for small distances around 3.5 μm, and shift parameter $0.15 \leq s \leq 0.25$. Precisely these set of parameters coincides with temporal stability, assessed by mapping the temporal oscillations of the central laser. Inspecting Figs. 4. (e) and (f), we assert that temporal stability is achieved for a simultaneous light localization. In this case, temporal stability reappears at larger distances for significant values of the shift parameter.

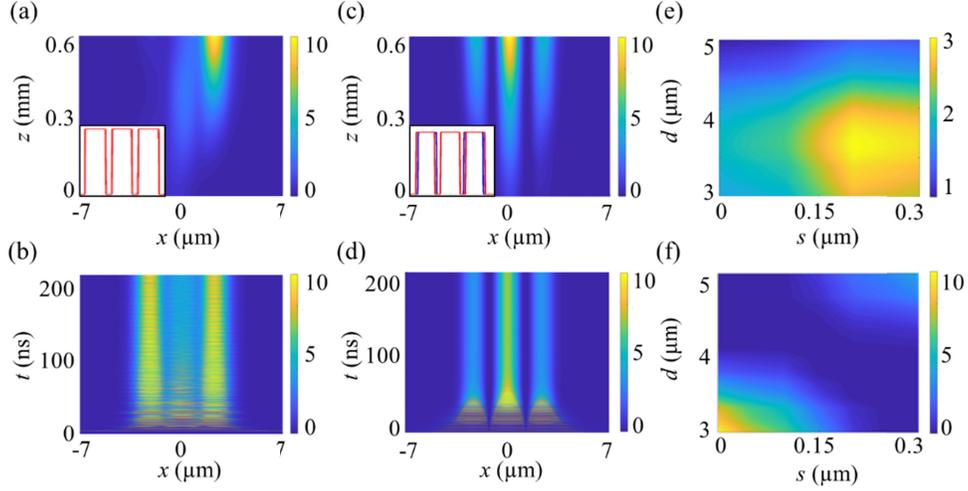

**Fig. 4**. (a)/(c) Snapshots of the spatial intensity distribution of the intensity for a triad of lasers with symmetric/inward-axisymmetric coupling after sufficient integration time and (b)/(d) temporal evolution of the intensity at $z = L$ for three lasers with symmetric/inward-axisymmetric coupling with $w = 2.5$ μm, separated a center-to-center distance $d = 3.0$ μm and for $s = 0.25$ μm. Insets shows the transverse pump profile $\Delta p$ (red curve) and the index profile $\Delta n$ (blue curve), and they coincide in (a). (e) Relative intensity and (f) amplitude of the temporal oscillations depending on the distance between lasers, $d$, and spatial shift, $s$. All the integration parameters are the same as in Fig. 2.

## 5. EEL array with axisymmetric inward coupling

The results presented in the previous section provide a proof of principle along with a comprehensive understanding of the asymmetric coupling of the proposed scheme. Next such results are extended to an EEL array or an EEL bar composed of many lasers.

The proposed scheme is depicted in Fig. 5. (c), the array is divided into two half-spaces with symmetric spatial shift between gain and index transverse profiles, arranged such that the index lays closer to the symmetry axis and the central laser holding no displacement. Thus, we expect light generated within the array to be directed inwards due to the asymmetric coupling.

We develop a simplified model, accounting for the longitudinal propagation, temporal evolution and coupling between lasers but disregarding diffraction , $\frac{\partial^2 A}{\partial x^2}$ , and carriers diffusion, $\nabla^2 N = 0$ , so considering a transverse mean field for each laser (no transversal modes). The simplified model is aimed at showing the concept on EEL bars, and to characterize the asymmetric coupling with a large number of lasers. The model describes the forward and backward field of every single laser of the array, namely $A_j^\pm$, as monomode lasers in the transverse direction, only coupled to the two neighboring lasers, as:

$$\frac{\partial A_j^\pm}{\partial z} = \sigma\left[(1-ih)N - (1+\alpha)\right]A_j^\pm + m^- A_{j-1}^\pm + m^+ A_{j+1}^\pm$$

$$\frac{\partial N_j}{\partial t} = \gamma\left(-N - (N-1)\left(|A^+|^2 + |A^-|^2\right) + p_0\right), \text{ for } j = -\frac{num-1}{2},\ldots,\frac{num-1}{2} \quad (2)$$

where, $m^{+/-}$ are complex numbers standing for the coupling parameter from the lasers at positions $j+1$ and $j-1$, respectively, and the spatial and temporal coordinates and all parameters are normalized as in eq. (1). In the simplest case, a periodic non-Hermitic potential in one-dimension may be approximated by a complex harmonic form: $V(x) = m_m\left[m_r \cos\left(\frac{2\pi}{d}x\right) + im_i \sin\left(\frac{2\pi}{d}x + \Phi\right)\right] = m_m\left[m^+ e^{+i\frac{2\pi}{d}x} + m^- e^{-i\frac{2\pi}{d}x}\right]$, where $d$ is the distance between lasers and $m_r$ and $m_i$ the amplitudes of the real and imaginary part of the non-Hermitian potential. Therefore, from eq. (2) we may express the coupling between neighboring laser as deriving from this simple harmonic complex potential as $m^\pm = m_m\left[\left(\frac{m_r \pm m_i}{2}\right)\cos\Phi + im_i\sin\Phi\right]$ being $m_m$ the coupling strength. In this model we can define different types of coupling between two lasers depending on the values of $m_r$, $m_i$ and the phase, $\Phi$. For $\Phi = \pm \pi/2$, the coupling is perfectly symmetric while for $\Phi = 0$, the coupling becomes PT-symmetric. In turn, and for simplicity we assume $m_r = m_i = 1$, the PT-symmetry breaking point, which entails no restriction but assuming the maximally asymmetric situation. Despite the relationship between parameters of spatial modulation in both models is nontrivial, we can infer a logarithmic relationship between $m_m$ and the laser distance $d$ as far as the coupling strength should decrease exponentially with the laser separation given by ($d$-$w$). This relationship is verified by direct comparison of the temporal instability in both models shown in Fig.5(b). The phase shift between the real and imaginary components of the coupling, $\pi/2$-$\Phi$ is proportional to the spatial shift $s$. The other parameters: $\sigma$, $h$, $\alpha$, $\gamma$ and $p_0$ are the same as in eq. (1).

The system of eq. (2) is integrated with boundary conditions given by the two mirrors of the Fabry-Perot cavity as for the complete model described by eq. (1). In order to compare the results of this simplified model to the complete model, we find equivalences between parameters comparing behaviors. We first determine the pump threshold $p_{th}$ and the onset of temporal instability, Hopf bifurcation $p_H$, for a single laser to locate equivalent pump values, see Fig. 5. (a). The temporal instability onset corresponds to smooth oscillations of small amplitude, just for a pump interval above the bifurcation, abruptly changing to a pulsed regime with short and bright pulses, starting from almost zero constant output power, as observed with the complete model in Fig. 1. (c). In order to provide a comparison between the coupling parameters of the simplified model, namely just $m_m$ and $\Phi$, coupling strength and phase shift, with the parameters of the full model, $d$ and $s$, distance between lasers and shift between the gain and index profiles, we consider two coupled lasers and map the amplitude of the temporal oscillations of the field amplitude of the laser to which the energy is transferred. Inspecting the comparison in Fig. 5. (b) we conclude that for a symmetric coupling, hence $\Phi = \pi/2$, instabilities arise for values of the coupling strength between lasers, $m_m$, above $10^{-6}$ equivalent to no asymmetric shift, $s = 0$ and distances $d$ larger than 5.0 μm in the full model. Moreover, for coupling strengths of $10^{-4}$ and a symmetric coupling the simplified model shows behaviors analogous to a distance between lasers $d = 3.0$-$3.5$ μm, and no asymmetric shift. As expected, the laser is temporally stable at the symmetry breaking point, hence for $\Phi = 0$, although the transition to stabilization is reached for much larger values of $\Phi$, about $\Phi \approx \pi/3$, which comparing the two models, turns out to be analogous to an asymmetric shift about $s \approx 0.3$ μm. Thus, both models agree that it is not necessary to reach totally asymmetric unidirectional couplings, but just a small shift, *s*, or, equivalently, a small phase shift (π/2 - *Φ*) to attain stabilization. Apart from achieving temporal stabilization the proposed scheme is intended to localize the output beam profile to enable the direct coupling of the light generated from the entire bar directly into an optical fiber, eventually avoiding any optical elements and minimizing losses as depicted in Fig. 5. (**c**).

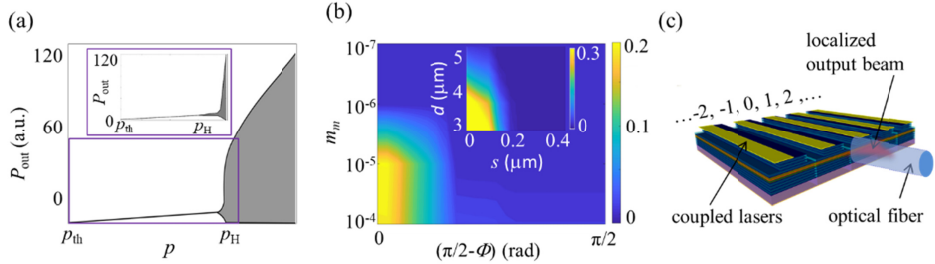

**Fig. 5.** (a) LI-curve for a single laser numerically obtained integrating the simplified model, described by eq. (2). The inset shows the LI-curve obtained by the complete model, eq. (1), for comparison. (b) Temporal instability map. Relative amplitude of the temporal oscillations (normalized to the average amplitude) for two coupled lasers obtained with the simplified model, depending on coupling magnitude, $m_m$, and phase shift, π/2-*Φ*, between coupling coefficients. The inset provides a comparison to the complete model, amplitude of the temporal instability depending on the transverse shift coefficient, s. All the integration parameters are the same as in Fig. 2. (c) Scheme of the array of coupled lasers directly coupled to an optical fiber. The integration parameters for the simplified model are the same as in Fig. 2. with $m_m = 5 \cdot 10^{-3}$ and *Φ* = 0.

To confirm the effect in a laser array we use both, the complete and the simplified model to numerically calculate the spatial redistribution of the generated light, the localization and the light enhancement at the central laser, for arrays of many lasers, schematically shown in the inset of Fig. 6. (b). We use equation (1) to evaluate the decrease of the beam quality factor $M^2$, for arrays up to 13 lasers, see Fig.6 (a), due to the stripping (black line) and the non-Hermitian potential effect (blue line) with respect a single broad laser with the same pumped area (orange line). We consider an array formed by *num* lasers and increase *num* up to 21 in the simplified model, for being a sensible number of lasers for an actual laser bar. Using equation (2), we assume *Φ* = 0, and consider two different coupling factors corresponding to a strong ($m_m = 10^{-3}$) and weak ($m_m = 10^{-5}$) coupling. For a strong coupling, as a consequence of the field localization induced by the asymmetric inward coupling, the output power of the central laser increases with the number of lasers, see Fig. 6. (a). Yet, for a small coupling the output intensity of the central remains almost constant. For a given laser array the intensity distribution, at each laser position, can be evaluated as a function of the coupling parameter, see Fig. 6. (b) showing the increase of the energy concentration towards the central laser rapidly increasing with the coupling parameter $m_m$ above a given threshold ($m_m = 10^{-3}$). The peak at the central laser is due to the addition of fields coming from both sides. While these results proof the working principle of the proposal we note that a more realistic approach indented to design an experiment should certainly include diffraction and inhomogeneous losses reducing the field localization.

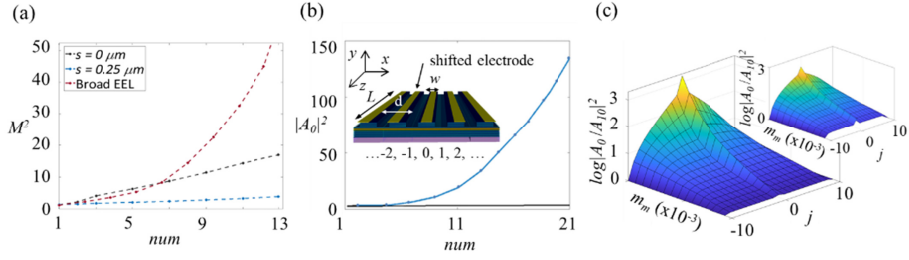

**Fig. 6.** (a) $M^2$ value dependence on the number of lasers for a symmetric array of lasers ($s = 0$ μm, black), a non-Hermitian array of lasers ($s = 0.25$ μm, blue) and a single broad EEL with equivalent pumped area (red). (b) Intensity of the central laser, $|A_0|^2$, for arrays of *num* lasers with inward axisymmetric coupling, for the simplified model, eq. (2), and for two different coupling strengths: $m_m = 10^{-3}$ (blue) and for $m_m = 10^{-5}$ (black). The inset displays the proposed scheme for *num* lasers, $j = -(num-1)/2…0…(num-1)/2$, with shifted gain and index profiles. (c) Emitted intensity distribution for an array made of *num* = 21 lasers as a function of the coupling strength, $m_m$. Intensity of laser $j$ relative to the laser at edge (j=10) in logarithmic scale. In the inset the equivalent emitted intensity distribution for the punctual laser model of eq. (3).

Finally, in order to provide a deeper analytic insight, we consider a stationary and punctual model, i.e. assuming null spatial and temporal derivatives in eq. (2) and just with inward axisymmetric coupling. The amplitude solution for the uncoupled lasers at each edge coincides with the standing alone laser: $|A|^2 = \dfrac{p_0 - (1+\alpha)}{\alpha}$, from which, we induce the amplitudes of all lasers of the array, from the edges to the central laser, resulting in:

$$\sigma\left[(1-ih)\frac{p_0 + |A_j|^2}{1+|A_j|^2} - (1+\alpha)\right]A_j + m_j^- A_{j+1} + m_j^+ A_{j-1} = 0, \text{ for } j = -\frac{num-1}{2},…,\frac{num-1}{2} \quad (3)$$

$m_{-j}^- = m_j^+ = 0$, $m_j^- = m_{-j}^+ = m_m$ for $j \geq 0$.

This simple estimation of the central laser intensity and energy distribution on the array really shows a good agreement with the numerical integration of the forward and backward fields in eq. (2) for the *num* asymmetrically coupled cavities. In spite of the simplicity of the punctual model, comparing Fig. 6. (a) and (b) with (c), we observe an analogous trend, yet in Fig. 6. (c) the pump is assumed to be just above the lasing threshold. Both results show a similar dependence on the number of lasers and coupling strength, and for the particular case of *num* = 21, both energy distribution patterns strongly change their profiles for the same critical value of $m_m$, about $m_m = 10^{-3}$.

We can simply deduce from eq. (3) that for a weak coupling strength $m_m$, the gain enhancement in laser $j$ given by the asymmetric coupling from the neighboring laser $j+1$ (or $j-1$) is small and the laser losses $\alpha$ limits the field amplitude $A_j$ remaining constant along the array, from the edge lasers to the central laser, with an intensity value about the standing alone laser $|A_j|^2 \approx \dfrac{p_0 - (1+\alpha)}{\alpha}$. In contrast, for large $m_m$ values, the asymmetric coupling strongly enhances gain, and the cascade effect from edges to the central laser is the cause of the energy localization with a sharp profile. Considering amplitude values much larger than the pump value, the multiplying factor becomes $\dfrac{A_{\pm|j|}}{A_{\pm|j+1|}} = \dfrac{m_m}{\sigma\alpha}$ from which we can infer the threshold coupling strength for a sharp profile in the energy distribution, $m_{mth} = \sigma\alpha = 10^{-3}$ for the considered parameters, in good agreement with numeric simulations.

## 6. Conclusions

In summary, we propose a physical mechanism for the temporal stabilization and localization of the emission of a coupled array of EELs or an EEL bar. The scheme is based on a non-Hermitian coupling between neighboring lasers with a global mirror-symmetric geometry. While the monomode emission of a single laser is assured by reducing its width, spatiotemporal instabilities may still arise from the coupling between lasers in an array. Such temporal instabilities are molded by a non-Hermitian coupling that may be simply introduced by a lateral shift between the pump and index profile, technically by a spatial shift between the individual laser stripe and corresponding electrode. Such asymmetric coupling, while temporally stabilizing the emission also redistributes and localizes energy close to the central symmetry axis.

The proposed stabilization scheme is analyzed by a complete spatiotemporal model, including transverse and longitudinal spatial degrees of freedom and the temporal evolution of the electric fields and carriers. We perform a comprehensive numerical analysis in terms of the design parameters, namely the distance between lasers and non-Hermitian shift observing regimes of simultaneous temporal stabilizations and light localization. In turn, the validity of the proposal is also demonstrated for an array with a large number of lasers using a simplified model where the transverse space is accounted by the coupling between neighboring laser cavities.

The proposed non-Hermitian architecture for stable EEL bars is demonstrated leading to a brighter output beam, yet the field concentration is expected to facilitate a direct coupling of these semiconductor lasers arrays to an optical fiber.